\documentclass[
reprint,
superscriptaddress,
showpacs,
preprintnumbers,
nofootinbib,
nobibnotes,
amsmath,
amssymb, 
aps,
prd,
floatfix
]{revtex4-1}

\usepackage[utf8]{inputenc}
\usepackage[normalem]{ulem}
\usepackage{graphicx}
\usepackage{dcolumn}
\usepackage[colorlinks=true,allcolors=purple]{hyperref}
\usepackage{url}
\usepackage{enumerate}

\usepackage{slashed,multirow,relsize,soul,feynmp-auto,tikz}
\usepackage{color}
\usepackage{mathrsfs} 
\usepackage{amsmath}
\usepackage{cancel}
\usepackage{bbold}
 \usepackage{mathrsfs}
\usepackage{braket}
\usepackage{physics}
\usepackage{multirow}
\usepackage[capitalize]{cleveref}
\usepackage{xspace}
\usepackage{fontawesome} 
\definecolor{blue-violet}{rgb}{0.33, 0.17, 0.89}

\definecolor{MH}{rgb}{0.0,0.6,9}
\definecolor{palatinate}{rgb}{0.494, 0.192, 0.482}
\definecolor{blue-violet}{rgb}{0.33, 0.17, 0.89}

\begin{document}

\title{Resonant Neutrino Flavor Conversion in the Atmosphere}

\author{Connor Sponsler}
\email{csponsler@college.harvard.edu}
\affiliation{Department of Physics \& Laboratory for Particle Physics and Cosmology, Harvard University, Cambridge, MA 02138, USA}

\author{Matheus Hostert}
\email{mhostert@g.harvard.edu}
\affiliation{Department of Physics \& Laboratory for Particle Physics and Cosmology, Harvard University, Cambridge, MA 02138, USA}

\author{Ivan Martinez-Soler}
\affiliation{Institute for Particle Physics Phenomenology, Department of Physics, Durham University, Durham DH1 3LE, U.K.}

\author{Carlos A. Argüelles}
\affiliation{Department of Physics \& Laboratory for Particle Physics and Cosmology, Harvard University, Cambridge, MA 02138, USA}

\date{\today}

\begin{abstract}
Neutrinos produced in the atmosphere traverse a column density of air before being detected at neutrino observatories like IceCube or KM3NeT.
In this work, we extend the neutrino flavor evolution in the \textsc{nuSQuIDS} code accounting for the varying height of neutrino production and the variable air density in the atmosphere.
These effects can lead to sizeable spectral distortions in standard neutrino oscillations and are crucial to accurately describe some new physics scenarios.
As an example, we study a model of quasi-sterile neutrinos that induce resonant flavor conversions at neutrino energies of $\mathcal{O}(300)\text{ MeV}$ in matter densities of $1 \text{ g/cm}^3$.
In atmospheric air densities, the same resonance is then realized at neutrino energies of $\mathcal{O}(300- 700)$~GeV.
We find that the new resonance can deplete the $\nu_\mu + \overline{\nu}_\mu$ flux at the IceCube Neutrino Observatory by as much as $10\%$ in the direction of the horizon.
\end{abstract}

\maketitle

\section{Introduction}
\label{sec:introduction}

Solar neutrinos undergo resonant flavor conversion inside the Sun thanks to the high-density and smoothly varying background matter.
The Mikheev-Smirnov-Wolfenstein (MSW) effect~\cite{Wolfenstein:1977ue,Mikheyev:1985zog}, as it became known, is now well-established and describes the flavor evolution of neutrinos in several environments, including supernovae, accelerator, and atmospheric neutrino sources.
This effect combined with vacuum oscillations and non-adiabatic matter effects successfully explains most current neutrino data and gave rise to the three-neutrino paradigm.
In this paradigm, the Standard Model (SM) is extended to include neutrino masses, of either Dirac or Majorana nature, such that at least two neutrino masses are non-vanishing with mass splittings measured by neutrino oscillation experiments, namely $\Delta m^2_{21}\simeq 7.5\times 10^{-5}$~eV$^2$ and $|\Delta m^2_{31}| \simeq 2.5 \times 10^{-3}$~eV$^2$~\cite{Esteban:2020cvm}.
The sign of $\Delta m^2_{31}$ is still unknown and will be measured using atmospheric~\cite{Arguelles:2022hrt}, reactor~\cite{Forero:2021lax,IceCube-Gen2:2019fet}, and accelerator neutrinos~\cite{Choubey:2022gzv}.
While this paradigm has successfully described most experimental data, it still does not preclude additional new physics from being found in the neutrino sector.

New interactions between neutrinos and ordinary matter can modify the matter potential neutrinos feel and lead to exotic flavor conversion.
The absence of such exotic conversions in data can be used to set strong bounds on new physics~\cite{IceCubeCollaboration:2021euf,IceCube:2022ubv}.
At the same, the anomalies observed by short-baseline neutrino experiments, like the apparent $\nu_\mu \to \nu_e$ conversion at the MiniBooNE~\cite{MiniBooNE:2020pnu} and $\overline{\nu}_\mu \to \overline{\nu}_e$ conversions at LSND~\cite{LSND:2001aii}, have motivated several explanations based on new matter potentials in the framework of effective field theories~\cite{Akhmedov:2010vy,Karagiorgi:2012kw,Kopp:2014fha,Esmaili:2018qzu}, light mediators coupled to the SM~\cite{Nelson:2007yq,Bramante:2011uu,Denton:2018dqq}, coupling to background fields~\cite{Fardon:2003eh,Kaplan:2004dq,Zurek:2004vd}, and in theories of modified dispersion relations~\cite{Pas:2005rb,Hollenberg:2009bq,Hollenberg:2009ws,Carena:2017qhd,Doring:2018cob}.
In this article, we turn to a recent proposal with quasi-sterile neutrinos: heavier neutrino states with no Weak charge but with sizeable interactions with ordinary matter~\cite{Alves:2022vgn} (akin to the proposals in Refs.~\cite{Nelson:2007yq,Bramante:2011uu,Esmaili:2018qzu,Denton:2018dqq}).
In these models, neutrinos can undergo resonant flavor conversion to quasi-sterile neutrinos, which subsequently mix with the electron flavor, thereby inducing $\nu_\mu$ disappearance as well as $\nu_\mu \to \nu_e$ appearance.

Our key insight is that if a new matter resonance is induced by protons, neutrons, or electrons at short-baseline experiments ($E_\nu \gtrsim 300$~MeV), then there would necessarily be a corresponding resonance for medium-energy ($E_\nu \gtrsim 0.5 - 1$~TeV) atmospheric neutrinos as these traverse the low-density air column in the atmosphere.
This effect can be searched for as an anomalous amount of $\nu_\mu$ disappearance at neutrino telescopes like IceCube and KM3NeT~\cite{KM3Net:2016zxf}, similar to existing searches for sterile neutrinos in up-going neutrino events~\cite{IceCube:2016rnb,IceCube:2020tka,IceCube:2020phf}.
Contrary to the parametric resonance induced by the Earth's core, crust, and mantle in the standard eV-sterile searches~\cite{Nunokawa:2003ep}, our scenario relies purely on the atmosphere and is most prominent for neutrinos coming from the direction of the horizon.
That such a resonance would be observable is, in fact, not trivial.
It requires that the resonance remains sufficiently narrow despite the smoothly varying air density and that the atmospheric neutrino production takes place at sufficiently large altitudes to allow the resonance to develop fully.
We demonstrate that both of these conditions are satisfied in the direction of the horizon at IceCube.

Another motivation for our work was improving the treatment of the atmosphere in publicly available codes for neutrino flavor evolution.
We focus here on the Neutrino Simple Quantum Integro-Differential Solver (\textsc{nuSQuIDS}) package~\cite{Arguelles:2021twb}, modeling the atmosphere as an extended spherical shell.
This provides a more accurate description of the path traveled by neutrinos than the usual assumption of an infinitely thin shell, shifting and smearing the oscillation probabilities in the three-neutrino paradigm.
These effects are significant for the oscillation of ordinary atmospheric neutrinos and become mandatory in new-physics scenarios such as the quasi-sterile neutrino model studied here.
The importance of the three-dimensional geometry of the atmosphere on the prediction of atmospheric neutrino fluxes has been long appreciated in the literature~\cite{Battistoni:1999at,Lipari:2000wu,Honda:2001xy}, especially at low energies with the Super-Kamiokande experiment~\cite{Super-Kamiokande:2005mbp,Super-Kamiokande:2017yvm}.
To our knowledge, however, this work is the first to include this effect in a publicly available code for neutrino flavor evolution.

\begin{figure}[t]
    \centering
    \includegraphics[width=0.48\textwidth]{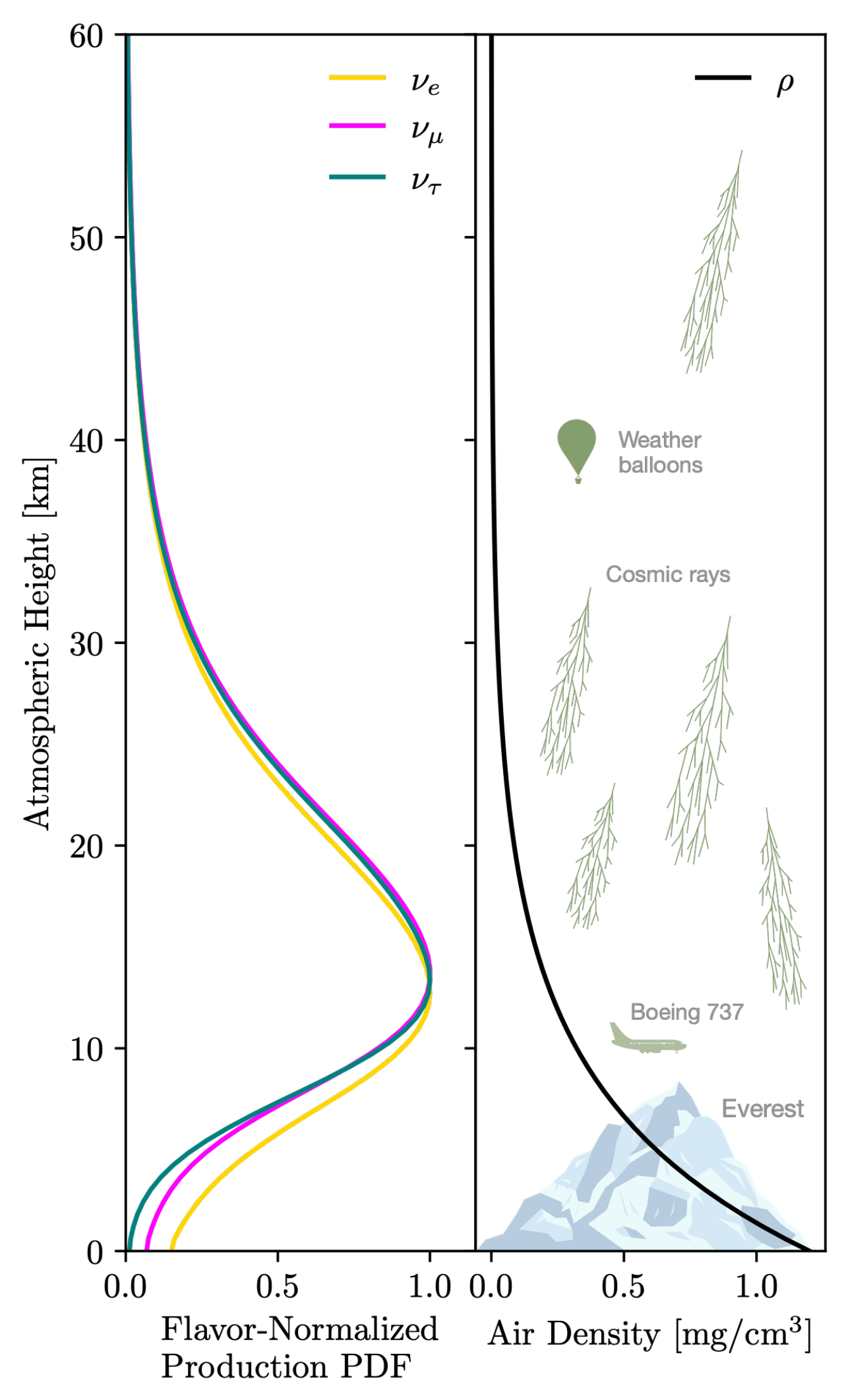}
    \caption{The flavor-normalized production of downgoing atmospheric neutrinos with $\cos\theta_z = 1$ (left) and the air density ($\rho$) of the atmosphere (right) as a function of the height at the South Pole in January.
    \label{fig:cartoon}}
\end{figure}

\begin{figure}[t]
    \centering
    \includegraphics[width=0.49\textwidth]{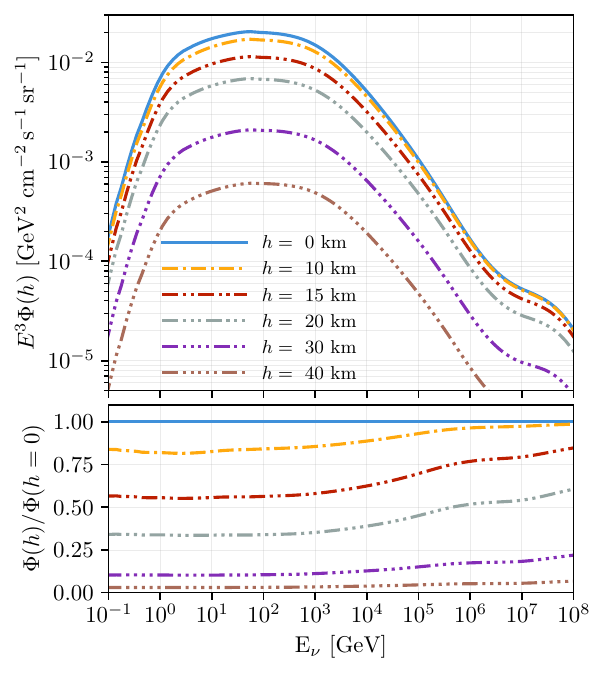}
    \caption{The top panels show the total atmospheric neutrino fluxes as a function of neutrino energy $E_\nu$ at different heights $h$. 
    The solid blue line corresponds to the total neutrino flux at the surface.
    The bottom panel shows the ratio between the flux at different heights and the surface.
    \label{fig:atmospheric_flux}}
\end{figure}

The rest of this article is divided as follows. 
In \cref{sec:production}, we describe neutrino production in the atmosphere and how it can be simulated with the Matrix Cascade Equations (\textsc{MCEq}).
We evaluate the impact of accurately describing the height dependence and the matter potential caused by the air density on standard oscillations in \cref{sec:standard_oscillation}.
We then discuss an application to new-physics models in \cref{sec:new_physics} and conclude in \cref{sec:conclusions}.

\section{Neutrino production in the atmosphere}
\label{sec:production}

\begin{figure*}[t]
    \centering
    \includegraphics[width=\textwidth]{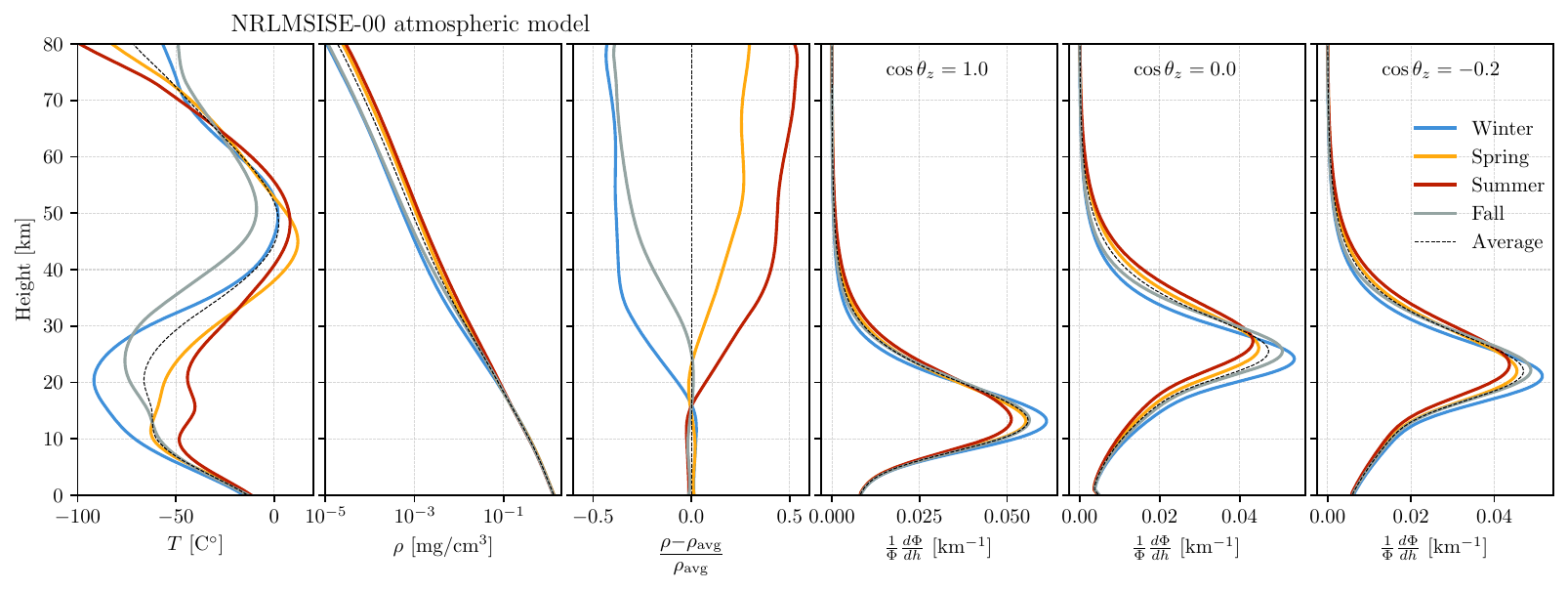}    
    \caption{From left to right, the panels show the temperature $T$, air density $\rho$, deviation from the average air density, and the normalized differential $\nu+\overline{\nu}$ flux $\frac{1}{\Phi}\frac{d\Phi}{dh}$ for a few indicative values of $\cos\theta_z$ as a function of atmospheric height during each of the Southern Hemisphere seasons.
    The average values for these quantities are shown as a dashed black line, where the average is weighted according to the total neutrino flux at each height.
    These plots correspond to the NRLMSISE-$00$ atmospheric model in MCEq.
 \label{fig:seasonal_var}}
\end{figure*}

The cosmic-ray flux measured at Earth spans over more than ten orders of magnitude in energy, from the GeV scale to energies above $\sim 100$~EeV.
For energies below the ``knee'' ($\sim 3$~PeV), the cosmic-ray flux decreases with energy as $\sim E^{-2.7}$.
The composition of this flux is mainly free protons ($\sim 80\%$) and bound nuclei ($\sim 20\%$)~\cite{Gaisser:2002jj,Dembinski:2017zsh}.
As cosmic rays propagate through the atmosphere, they collide with atmospheric nuclei, generating a secondary flux of mesons, predominantly composed of pions and kaons.
After propagating through the atmosphere, pions predominantly decay into a muon and a neutrino flux.
Other decay modes are also possible for kaons, involving charge pions, neutral pions, electrons, and electron neutrinos.
At energies below $\sim 1$~GeV, muons decay before reaching Earth's surface, contributing to the neutrino flux.
Pion decay dominates the neutrino flux below $\sim 100$~GeV, while kaon decay becomes more significant above that energy.
The prompt decay of the mesons creates a flux of $\nu_{e}$ and $\nu_{\mu}$ following the cosmic-ray spectrum. 
However, as the mesons interact with Earth's atmosphere and begin losing energy, the neutrino spectral index softens by nearly one unit~\cite{Gaisser:2016uoy}.

\Cref{fig:cartoon} shows the production height distribution of neutrinos and the air density as a function of the height above the Earth's surface.
Most atmospheric neutrinos are produced at a height between 10~km and 20~km due to the exponentially increasing air density, hadronic cross sections, and the lifetime of the parent mesons.
The energy dependence of the total atmospheric neutrino flux is shown in~\Cref{fig:atmospheric_flux}.
At very high energies, above $10^6$~GeV, the atmospheric neutrino flux is primarily comprised of byproducts of charmed mesons, which also contribute to a $\nu_{\tau}$ flux in the atmosphere~\cite{Enberg:2008te,Gauld:2015kvh}. 

The propagation of mesons through the atmosphere also leads to anisotropies in the neutrino flux.
For instance, the longer paths that mesons must travel in horizontal directions before reaching the surface of the Earth enhance the neutrino flux along those directions.
Additionally, the interaction of the cosmic rays and meson flux with the Earth's magnetic field also induces asymmetries in the neutrino flux, such as the East-West asymmetry~\cite{Super-Kamiokande:1999mpf,Super-Kamiokande:2015qek} or the enhancement of the neutrino flux over the anti-neutrino, as the rescattering on meson propagation due to geomagnetic effects becomes less significant at higher energies, allowing neutrino production to dominate.

\begin{figure*}[t]
    \centering
    \includegraphics[width=0.324\textwidth]{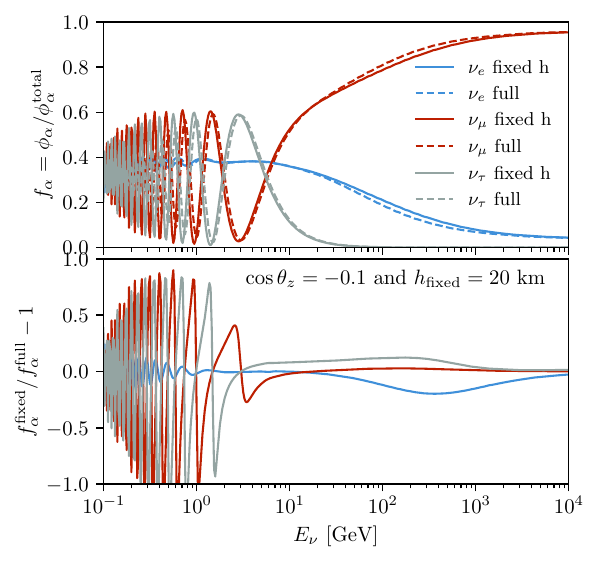}
    \includegraphics[width=0.324\textwidth]{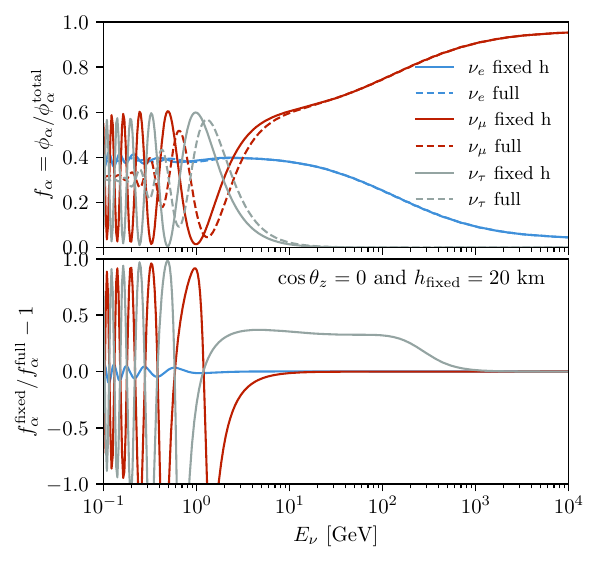}
    \includegraphics[width=0.324\textwidth]{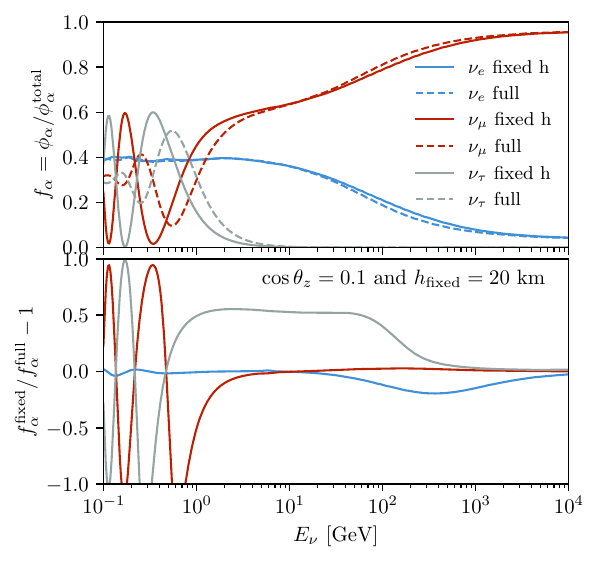}
    \caption{
    The flavor ratio of the neutrino flux as a function of neutrino energy.
    The top panels show the fixed-height (infinitely thin atmospheric shell) approximation as a solid line and the full description of the atmospheric shell source as a dashed line for each flavor.
    The relative error between each of the two methods, $f^{\rm fixed}_{\alpha}/f^{\rm full}_\alpha - 1$, is shown below each panel.
    From left to right, we show the case for $\cos\theta_z = -0.1$, $\cos\theta_z = 0$, and $\cos\theta_z = 0.1$.
    \label{fig:ratio_of_ratios_1D}}
\end{figure*}

\begin{figure*}[t]
    \centering
    \includegraphics[width=0.324\textwidth]{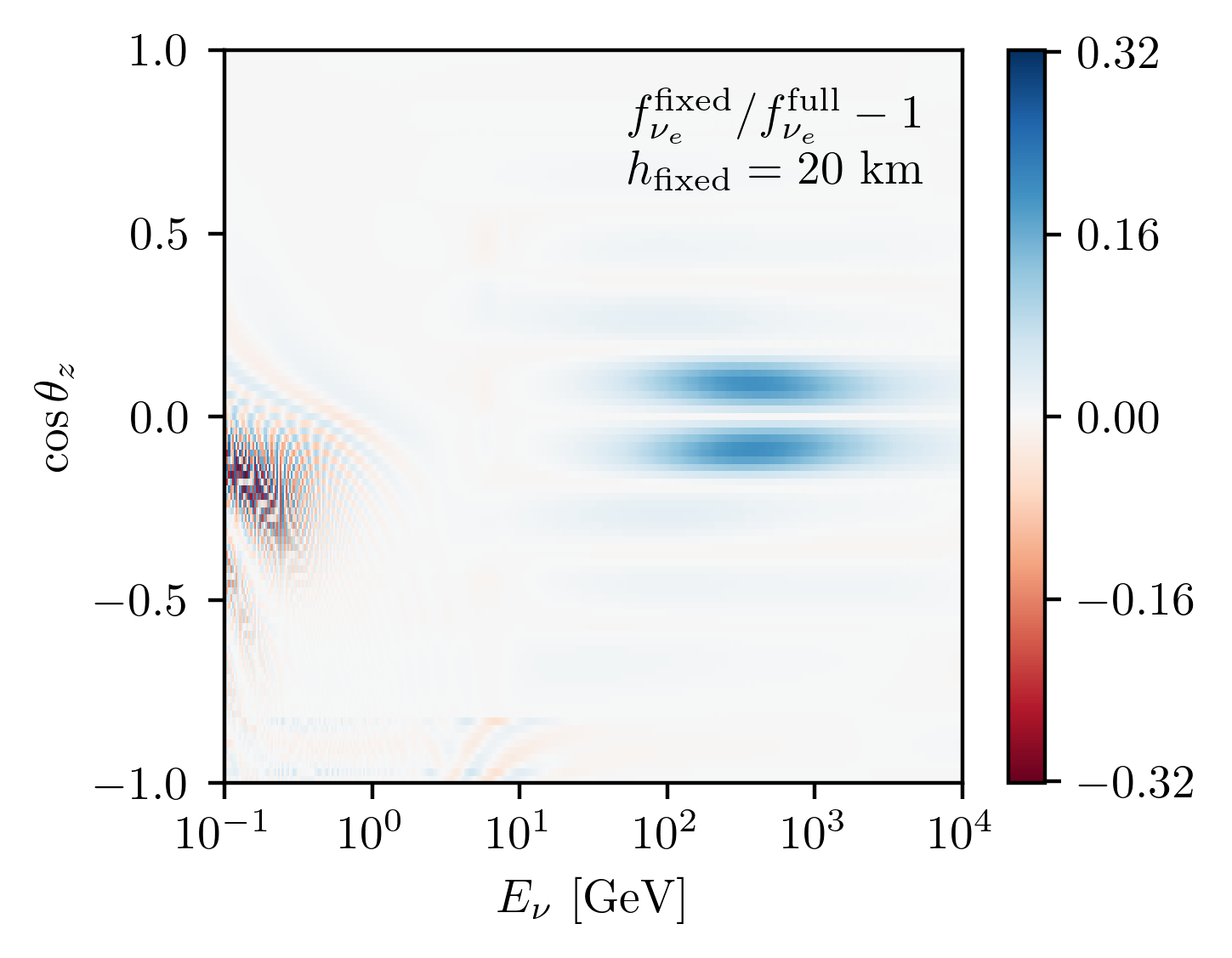}
    \includegraphics[width=0.324\textwidth]{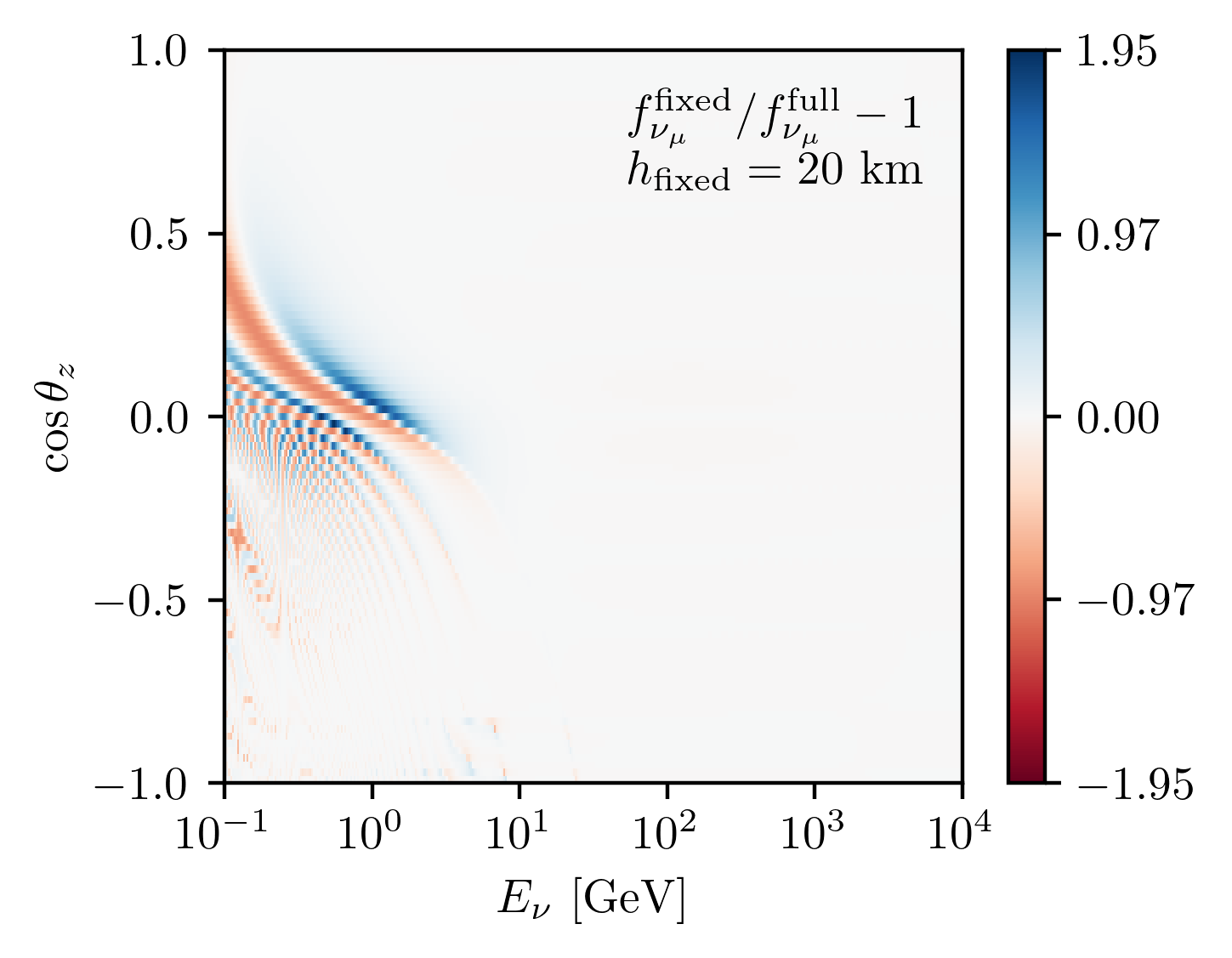}
    \includegraphics[width=0.324\textwidth]{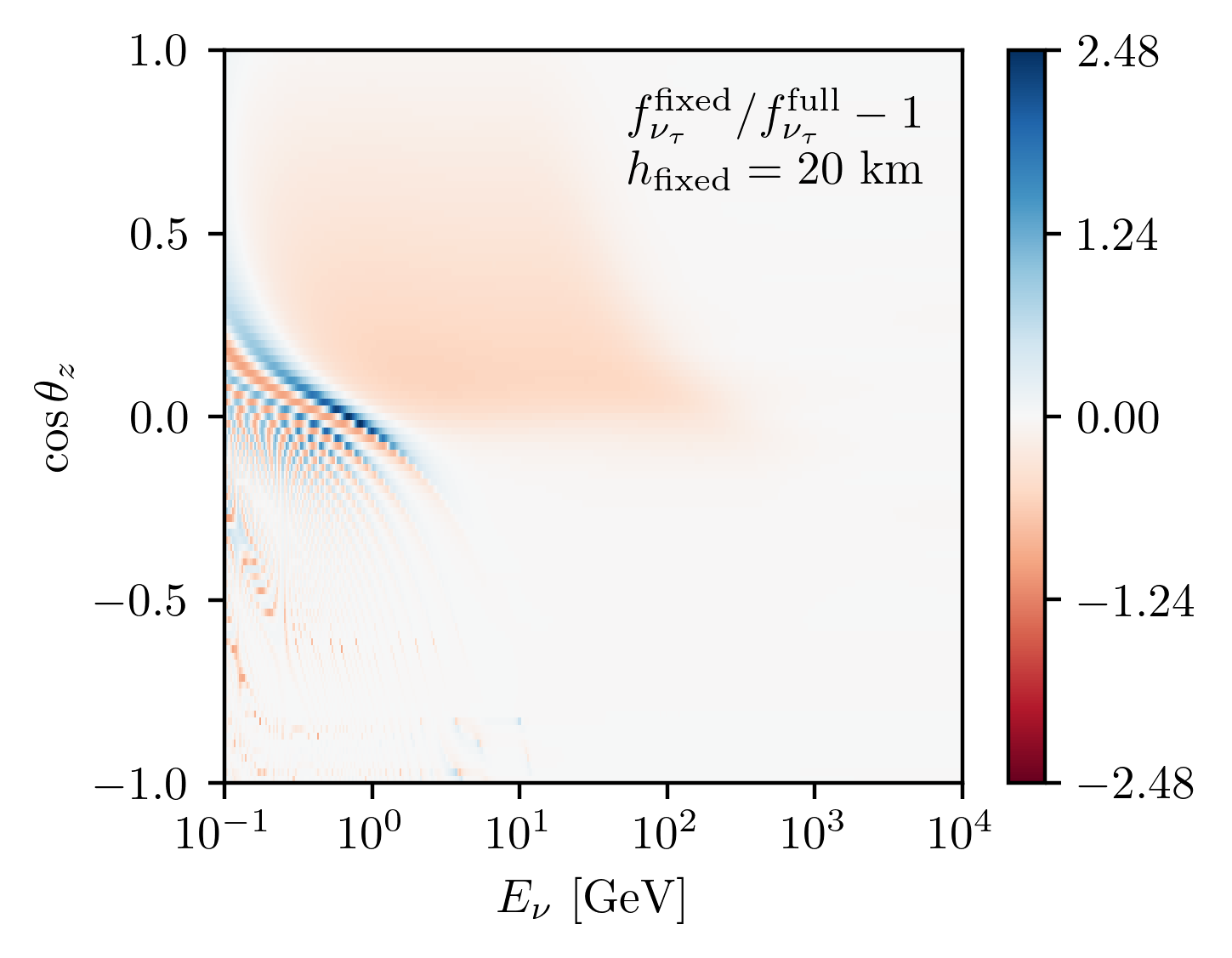}
    \caption{
    Comparison between the fixed-height approach and the full description of the atmospheric shell source.
    Each panel shows the relative error of the flavor ratios between the two approaches, $f^{\rm fixed}_{\alpha}/f^{\rm full}_\alpha - 1$, for different neutrino flavors $\alpha = e,\mu, \tau$ in the left, center, right panels, respectively.
    \label{fig:ratio_of_ratios_2D}}
\end{figure*}

In this work, we obtain the neutrino flux from \textsc{MCEq}~\cite{Fedynitch:2012fs}. 
This code describes the propagation and evolution of cosmic-ray showers using integro-differential cascade equations that describe the number density of particles in intervals of the shower's depth and energy, accounting for particle interactions with the medium, decay, and energy losses.
Because \textsc{MCEq} tracks the number density rather than individual particle histories, it is computationally faster and more tractable than full Monte-Carlo methods employed in larger open-source codes like \textsc{GEANT4}~\cite{GEANT4:2002zbu}, \textsc{FLUKA}~\cite{Ferrari:2005zk}, \textsc{CORSIKA}~\cite{Heck:1998vt}, or private codes like \textsc{HKKMS}~\cite{Honda:2004yz}.

The nucleon cosmic-ray flux at the top of the atmosphere is taken to be the Hillas-Gaisser-2012 model~\cite{Gaisser:2011klf}, which sets the initial conditions for the cascade equations. 
\textsc{MCEq} also requires an interaction model to describe hadronic interactions such as $p-p$, $p-\pi$, and $K-p$ interactions during the development of extensive air showers.
We follow the model \textsc{SIBYLL-2.3c}~\cite{Riehn:2017mfm}, based on the associated simulation code~\cite{Fletcher:1994bd}.
The temperature and density profile of the atmosphere at the South Pole is set by the \textsc{MSIS00\_IC} class in MCEq, which is based on NASA's \textsc{NRLMSISE-00} empirical atmosphere model~\cite{https://doi.org/10.1029/2002JA009430}, release 20041227.
This model is informed by a variety of measurements of the atmosphere before 2001.

\Cref{fig:seasonal_var} shows the seasonal variation of the atmosphere's neutrino flux, density, and temperature as a function of the height (or altitude).
These are known to be particularly important at the South Pole due to the temperature and climate variations in the Troposphere and Stratosphere.
Below $1$~TeV, the largest variation is between Summer and Winter, with the neutrino flux being less than $15\%$ smaller during the Winter than in the Summer.
Seasonal variations of the neutrino flux can be accurately modeled with the use of direct measurements of the atmospheric temperature, as recently demonstrated by IceCube~\cite{IceCube:2023qem} using data from the Atmospheric Infrared Sounder (AIRS) instrument aboard NASA’s AQUA satellite.

\section{Improved description of atmospheric neutrinos in \textsc{nuSQuIDS}}
\label{sec:standard_oscillation}

After the neutrino is produced in the atmosphere, it must travel from $\sim 15$~km for downgoing trajectories to $\gtrsim 12,000$~km for upgoing trajectories before reaching the detector.
In the three-neutrino mixing scenario, neutrino oscillation is described as the interference between three massive states, $\ket{\nu_{\alpha}} = \sum_{i} U_{\alpha i}\ket{\nu_{i}}$, where $U_{\alpha i}$ are the elements of the Pontecorvo-Maki-Nakagawa-Sakata (PMNS) matrix.
In this scenario, neutrino evolution is described by two mass-squared differences, $|\Delta m^2_{31}| \sim 2.5\times 10^{-3}~\text{eV}^2$ and $\Delta m^2_{21} \sim 7.5\times 10^{-5}~\text{eV}^2$, the three mixing angles ($\theta_{12}$, $\theta_{13}$ and $\theta_{23}$) and the CP-violation phase ($\delta_{\rm CP}$).
For energies above the GeV scale, atmospheric neutrinos undergo a flavor oscillation for trajectories crossing the Earth.
Those flavor oscillations are mainly described by $\Delta m^2_{31}$ and $\sin\theta_{23}$~\cite{Arguelles:2022hrt,Olavarrieta:2024eaq}. 

The neutrino oscillation probability depends, in addition to the oscillation parameters, on the distance traveled by the neutrino.
To accurately describe the flavor oscillation of atmospheric neutrinos, it is necessary to consider the different production altitudes where the neutrinos are generated in the atmosphere, as shown in~\Cref{fig:cartoon}, and the phase associated with it.
Compared to the case where it is assumed that the neutrinos are produced at a fixed altitude or at the Earth's surface (see~\Cref{fig:ratio_of_ratios_1D,fig:ratio_of_ratios_2D}), neutrino production at different altitudes introduces an averaging effect on the oscillation probability.
This becomes relevant when the oscillation length is comparable to the distance traveled by the neutrinos through the atmosphere. 
In the case of trajectories from down-going ($\cos\theta_{z} =1$) to horizontal ($\cos\theta_{z} =0$) directions, these effects are significant for neutrinos with energies below $E_{\nu} \leq 1$~GeV.

In~\Cref{fig:ratio_of_ratios_1D}, we compare the flavor ratios of the neutrino flux with an extended spherical shell and an infinitely thin spherical shell atmosphere.
\Cref{fig:ratio_of_ratios_2D} shows the same for all zenith angles and neutrino energies of interest.
The two approaches correspond to fixing the neutrino production to a given height ($h = 20$~km, in this case) and varying the neutrino production height according to the atmospheric neutrino model in \textsc{MCEq}.
The total neutrino flux at the surface is the same in both cases, but the path traveled by neutrinos and the atmospheric matter potential it experiences will differ.
The choice of the fixed height in the first approach is arbitrary and is usually set between $15 - 30$~km.
Changing this value in \Cref{fig:ratio_of_ratios_1D,fig:ratio_of_ratios_2D} shifts the oscillation features to higher or lower energies, but it cannot resolve the observed discrepancy between the two approaches for all angles and energies.
We also note that the fast oscillations at low energies will be averaged out by the finite detector resolution in energy and angle.
Therefore, the rapidly changing discrepancies at low energies will not be important for any realistic analysis.
However, the discrepancy can be of the order of $10\%$ at the $\nu_\mu \to \nu_e$ oscillation maximum.
We find very similar conclusions in the case of antineutrinos. 

\section{Atmospheric Resonance from new physics}
\label{sec:new_physics}

\begin{figure*}[t]
    \centering    \includegraphics[width=0.49\textwidth]{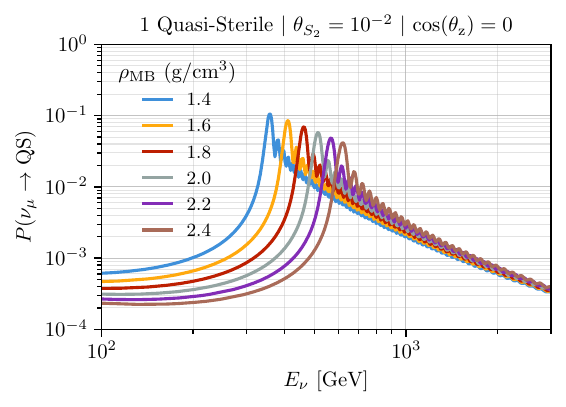}
    \includegraphics[width=0.49\textwidth]{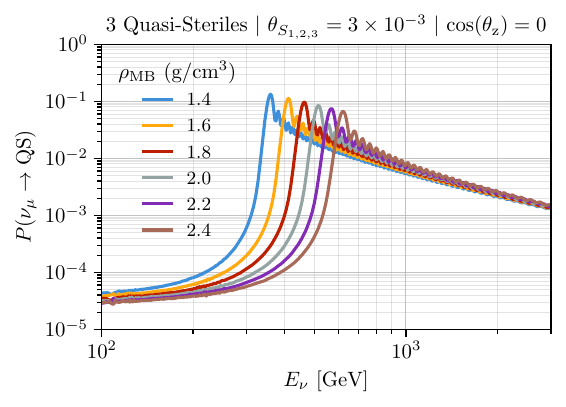}
    \caption{The muon neutrino disappearance probability in a model with a single quasi-sterile neutrino (left) and a model of three quasi-sterile neutrinos (right) in the direction of the horizon at IceCube.
    We fix the resonance energies and mass splitting as indicated in \cref{eq:theta_singleQS,eq:theta_threeQS}, corresponding to $E_{2}^{\rm res} = 300$~MeV at MiniBooNE.
    \label{fig:prob_benchmark}}
\end{figure*}

We now discuss an application of our calculations to new physics searches with neutrino telescopes. 
We study a scenario where neutrinos can undergo MSW-like resonant flavor conversion in the atmosphere due to a new matter potential for (quasi-)sterile neutrinos. 
Because the air density varies from the upper-most region of the atmosphere to Earth's surface, the resonance condition is satisfied at different neutrino energies.
Schematically, the resonant condition is given by $E_{\rm res} = \cos{2\theta} \times \Delta m^2/2V$, where $V$ is the matter potential caused by background fermions and $\Delta m^2$ is the mass-squared splitting between the new state and the light neutrinos.
The resonant energy is then inversely proportional to the background density.
This is the central point of our discussion: if accelerator neutrino experiments with $\mathcal{O}(100)$~MeV neutrinos were to observe a new matter-induced resonance in Earth's crust, then neutrino telescopes should be able to investigate this effect by searching for a corresponding resonance in $\mathcal{O}(300- 700)$~GeV neutrinos of atmospheric origin.

We are particularly interested in resonant flavor conversions that result in $\nu_\mu \to \nu_e$ appearance at accelerator experiments as well as $\nu_\mu$ and $\nu_e$ disappearance for neutrinos coming from the direction of the horizon at IceCube and other neutrino telescopes.
Discovery of the atmosphere-induced resonance would point to a new force between neutrinos and ordinary matter.
However, if short-baseline experiments were to confirm the existence of a resonance that does not show up in atmospheric neutrino experiments, one could then conclude that the new matter potential is sourced by something other than protons, neutrons, or electrons, such as by a dark matter background or by more exotic effects in neutrino propagation.

\subsection{Quasi-steriles neutrinos}

For concreteness, we will focus on the model proposed in Ref.~\cite{Alves:2022vgn}.
The model extends the three-neutrino paradigm by two families of neutral fermions: quasi-sterile neutrinos $(S_1, S_2, S_3)$ and sterile neutrinos $\nu_s$. 
The latter are fully sterile and do not feel any interactions with ordinary matter, while the former, as the name suggests, are subject to new interactions.
Active, sterile, and quasi-sterile neutrinos are coupled through mixing, giving rise to new oscillation frequencies and new matter effects in the flavor evolution of active neutrinos.
Ref.~\cite{Alves:2022vgn} introduced the model to explain the MiniBooNE low-energy excess with a combination of resonant flavor conversions driven by quasi-sterile neutrinos and vacuum short-baseline oscillations driven by a $\mathcal{O}(1)$~eV sterile neutrino.
As we are only interested in the effects of matter-induced flavor conversions on the atmosphere, we drop the sterile neutrino from our discussion.
This type of neutrino would predict a resonance for larger densities (or smaller energies) due to the SM neutral current potential~\cite{Nunokawa:2003ep}.
Instead, we focus exclusively on the quasi-sterile sector.
Like the authors, we remain agnostic about the UV completion behind the new matter potential and work with a generic parameterization of new interactions mediated by a new $U(1)^\prime$ vector boson (see, e.g., Refs.~\cite{Nelson:2007yq,Bramante:2011uu,Pospelov:2011ha,Pospelov:2012gm,Harnik:2012ni,Kopp:2014fha,Tabrizi:2015bba,Esmaili:2018qzu,Denton:2018dqq} for similar applications of such models).

The phenomenological Lagrangian of the model is given by
\begin{equation}
    \mathscr{L} \supset - A_\mu^\prime g_X \left( J^\mu_{\rm matter} + J^\mu_\nu \right),
\end{equation}
where  $A_\mu^\prime$ is a new vector boson responsible for the interaction of neutral leptons with ordinary matter, and the currents are given by 
\begin{align}\label{eq:current_nu}
    J^\mu_{\rm matter} = Q_p \overline{p} \gamma^\mu p + Q_n \overline{n} \gamma^\mu n + Q_e \overline{e} \gamma^\mu e,
    \\
    \label{eq:current_S}
    J^\mu_\nu = \sum_{i = 1}^{3} \left( Q_{\nu_i} \overline{\nu_i} \gamma^\mu P_L \nu_i + Q_{S_i} \overline{S_i} \gamma^\mu P_L S_i \right),
\end{align}
where $Q_f$ stands for the charge of the fermion $f$ under the $U(1)^\prime$ symmetry.
Ref.~\cite{Alves:2022vgn} parameterized the mixing between neutrinos and quasi-sterile neutrinos in an intermediate basis that would be identified with the neutrino mass basis in the absence of quasi-steriles.
Because quasi-steriles mix very weakly with SM neutrinos, the intermediate $\nu_i, S_i$ basis is a small perturbation away from the true mass basis $\nu_i^{+}, \nu_i^{-}$, where $i = \{1,2,3\}$.
Each single quasi-sterile neutrino $S_i$ mixes exclusively with a $\nu_i$ state, such that
\begin{align}
    \ket{\nu^-_i} &\simeq - \sin{\theta_i} \ket{S_i} + \cos{\theta_i} \left(\sum^{ e,\mu,\tau}_{\alpha} U_{\alpha i}^* \ket{\nu_\alpha}\right),
    \\
    \ket{\nu^+_i} &\simeq \cos{\theta_i} \ket{S_i} + \sin{\theta_i} \left(\sum^{ e,\mu,\tau}_{\alpha} U_{\alpha i}^* \ket{\nu_\alpha}\right),
\end{align}
where $\theta_{i} \ll 1$, all phases have been set to zero, and $U_{\alpha i}$ are the PMNS mixing matrix elements.

Under the assumption of isoscalar and neutral matter,\footnote{The isoscalar assumption for the atmosphere breaks down only at the sub-percent level due to the small concentration of inert argon gas, water vapor, and trace amounts of other isotopes.} the potential difference between SM neutrinos and quasi-sterile neutrinos is given by
\begin{align}
    \Delta V_i &= 
   V^{S_i}_{A^\prime} - V^{\nu_i}_{A^\prime}  - V^{\nu_i}_{\rm SM} \simeq - Q \frac{g_X^2}{2 m_{A^\prime}^2}n_e,
\end{align}
where $Q = (Q_n + Q_p + Q_e) (Q_{S_i}-Q_{\nu_i})$, $n_e$ is the electron number density in the material. The SM neutral-current and charged-current potential ($V^{\nu_i}_{\rm SM}$) is negligible compared to the one sourced by the new force but is included in the full Hamiltonian for the evolution of the active flavors.
We assume $Q = +1$, such that the potential difference is negative for {neutrinos} and positive for {antineutrinos}.

The condition for resonant \emph{neutrino} flavor conversion in matter is schematically given by,
\begin{equation}
    2E_i^{\rm res} |\Delta V_i| = \delta M^2_i \cos{2\theta_{S_i}},
\end{equation}
where $\delta M^2_{i} = m_{\nu^+_i}^2 - m_{\nu^-_i}^2> 0$ is the mass-squared splitting between the quasi-sterile and the corresponding light neutrino state.
For a new resonance at $E_i^{\rm res} \sim 300$~MeV, we consider the following benchmark values
\begin{equation}
\frac{E_{i}^{\rm res}}{300 \text{ MeV}} \sim Q
\left(\frac{m_{A^\prime}/g_X}{100 \text{ MeV}}\right)^2 \left( \frac{2.2 \text{ g/cm}^3}{\rho}\right)\left(\frac{\delta M^2_i}{230\text{ eV}^2}\right),
\end{equation}
where $\cos 2\theta_{S_i} \simeq 1$ and $\rho$ is the typical density of soil between the Booster neutrino target and the MiniBooNE detector.
The matter potential in that case is given by $\Delta V \sim 3.8\times 10^{-7}$~eV at MiniBooNE and $\Delta V \sim 2.3\times 10^{-10}$~eV at the surface of the Earth.
As expected, the new force must be significantly stronger than the Weak interaction to generate a sufficiently large potential at short baselines.

The largest sources of uncertainty in predicting the exact location of the resonance in the atmosphere are the choice for i) the resonant energy at MiniBooNE, which may vary between $E_\nu^{\rm res} = 200 - 400$~MeV and still explain the low-energy excess, and ii) the assumed value for the density $\rho$, which depends on the exact composition and moisture of the soil.
The soil at the MiniBooNE location comprises glacial till, a mixture of sand, silt, and clay, on top of a bedrock layer at about 60 ft depth~\cite{Awschalom:1969uc,Bauer:1996rp,MiniBooNE:2001sbk}.
The target and the detector are comfortably above the bedrock layer, so the relevant soil for neutrino propagation is the shallower, lower-density component.
With a survey of the literature, we determined that dry and wet glacial till have densities varying from 2.1 to 2.4 g/cm$^3$, while silty clay has typical densities of $1.4$ to $1.8$ g/cm$^3$, to be compared with silty gravel and bedrock that have densities exceeding $2.4$ and $2.6$ g/cm$^3$, respectively.
Given that we do not have access to the exact makeup of the soil and to density measurements that may have been carried out at the site, we follow Ref.~\cite{Alves:2022vgn} and quote our results for various density values, varying $\rho$ from $1.4$ to $2.4$~g/cm$^3$.
As a benchmark, we choose $2.2$~g/cm$^3$, close to the value of $2.25$ quoted in Ref.~\cite{Bauer:1996rp}.

\begin{figure*}[t]

    \centering
    \includegraphics[width=0.49\textwidth]{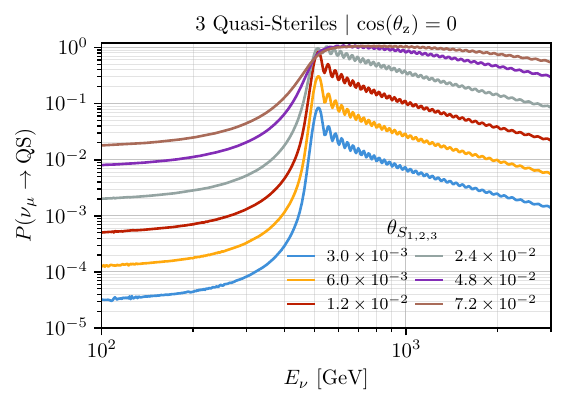}
    \includegraphics[width=0.49\textwidth]{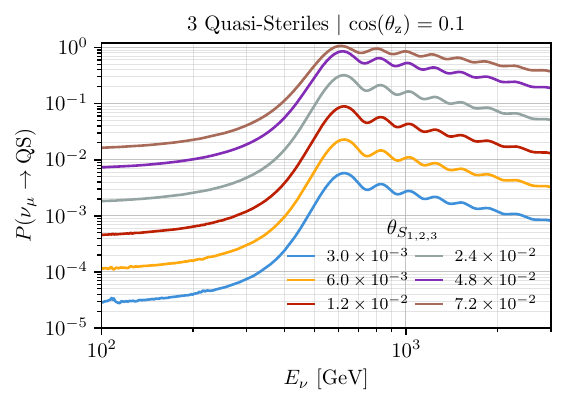}
    \caption{The probability for muon neutrino disappearance at IceCube for neutrinos from the horizon (left) and from  $\cos(\theta_z) = 0$ (right).
    Each curve corresponds to a different choice for the mixing angles $\theta_{S_{1,2,3}}$.
    We fix $E_{2}^{\rm res} = 300$~MeV and $\rho_{\rm MB} = 2.2$~g/cm$^3$.
    \label{fig:prob_diffmixing}}
\end{figure*}

The form of the currents in \cref{eq:current_nu,eq:current_S} is strongly constrained by the requirement of gauge invariance.
It narrows the choices to a $B-L$, electromagnetic, leptophilic $L_e - L_{\mu, \tau}$ current, or combinations thereof.
The coupling of $A^\prime_\mu$ to the electromagnetic current, e.g., generated by kinetic mixing with the photon, does not lead to a potential for quasi-steriles due to the neutrality of matter.
All remaining possibilities are strongly constrained by neutrino data from neutrino-electron and neutrino-nucleus elastic scattering at the MeV scale and above, neutrino oscillation experiments~\cite{Wise:2018rnb,Coloma:2020gfv}, cosmology, fifth-force experiments, and astrophysics.
Constraints on the mediator with large momentum exchange push the model towards small values of $m_{A^\prime}$, where bounds from astrophysics and cosmology dominate.
Additional model building may assuage the strong requirements of current conservation and will be required to reconcile the couplings required here with cosmology and astrophysical limits (see, e.g., Ref.~\cite{Nelson:2007yq} for an effort in this direction).

Long-range forces between neutrinos and background matter can also generate the matter potential. 
For instance, when $m_{A^\prime} \ll 1 \times 10^{-7}$~eV, the range of the new force becomes longer than a meter, allowing significantly more material to contribute to the matter potential of neutrinos.
For even longer ranges, the distinction between the matter potential at accelerator and atmospheric neutrino experiments becomes less well-defined, and even entire astrophysical bodies would contribute~\cite{Joshipura:2003jh,Grifols:2003gy,Heeck:2010pg,Davoudiasl:2011sz,Wise:2018rnb}.
Another possibility includes a potential sourced by a dark matter background~\cite{Denton:2018dqq,Smirnov:2019cae,Babu:2019iml,Davoudiasl:2023uiq}.
In that case, the density profile of dark matter particles in the crust and atmosphere would be similar, avoiding the argument presented here.
We also note that if this dark matter background was enhanced in the vicinity of the Earth due to a ``traffic-jam" effect (see, e.g., the models in Refs.~\cite{Pospelov:2020ktu,McKeen:2022poo}), the resonance may be visible at a different energy in the atmosphere, where the dark matter background density can be either smaller or larger than the one in the rock.
In all such scenarios, the value of the resonant energy for neutrinos in different environments is significantly more model-dependent.
In addition, the interaction strengths between quasi-sterile and dark matter should be significantly larger than those between quasi-sterile and ordinary matter, which is also constrained by cosmology.
We leave an exploration of the different interpretations of the matter potential to future literature and proceed with the understanding that for potentials sourced by ordinary matter, there must be additional model complications to avoid existing constraints on new forces~\footnote{This includes overcoming the challenge to explain the lack of a new resonant electron neutrino disappearance inside the Sun~\cite{Alves:2022vgn}.}.

\subsection{Muon neutrino disappearance}

The quasi-sterile model was proposed to explain hints of $\nu_\mu \to \nu_e$ appearance, but it leads to corresponding $\nu_{e,\mu,\tau}$ disappearance effects, thanks to the mixing of the quasi-steriles with all flavors.
Therefore, the most promising channel to search for the atmospheric resonance is $\nu_\mu$ disappearance as that is the largest atmospheric flux.
The model prediction will correspond to a sharp deficit of events in the $E_\nu$ vs $\cos\theta_z$ spectrum of atmospheric neutrinos.
As we will see, the size of the mixing angles $\theta_{S_{1,2,3}}$ are crucial to determining the observability of this effect at IceCube.
In what follows, we discuss the general features of the flavor evolution and present our numerical results. 

The flavor evolution of neutrinos is tracked fully numerically by \textsc{nuSQUiDS} by time evolving the $6\times6$ Hamiltonian in matter from a large height of $80$~km above the Earth's surface all the way to the South Pole, accounting for the height-dependent neutrino fluxes and assuming a perfectly spherical Earth. 
For simplicity, we track neutrinos to a single surface point on top of IceCube, neglecting the detector's ice overburden of about $1.5$~km.
Propagation through the Earth is computationally costly for upgoing neutrinos due to the fast phase evolution caused by the quasi-sterile potential, but we have checked that the oscillation probabilities are approximately symmetric for $-0.2 < \cos \theta_z < 0.2$ in the energy range of interest, $200 \text{ GeV}< E_\nu < 1 \text{ TeV}$.
Our results are quoted for the neutrino flux in July, but we find that the seasonal variation has a negligible impact on the disappearance probabilities.

In the quasi-sterile neutrino model, the flavor evolution of atmospheric neutrinos is dictated by length scales that are rapidly changing and of similar size.
For instance, for the resonance to fully develop within the atmosphere, neutrinos must acquire a full phase difference due to the matter potential, $\Delta \phi = \int \dd t |\Delta V| = 2\pi$.
The refractive length $\ell_r$ indicates the baseline at which this condition is met.
For quasi-sterile neutrinos,
\begin{equation}
    \ell_{r} = \frac{2 \pi}{|\Delta V|} \sim 6 \text{ km} \times \left(\frac{1 \text{ mg/cm}^3}{\rho}\right),
\end{equation}
which depends on the local air density.
A crude approximation to the air density profile is given by
\begin{equation}
    \rho \sim \rho_0 \times \exp\left(-{h/h_0}\right),
\end{equation}
with $h$ the height above the surface, $\rho_0 = 1.3 \text{ mg/cm}^3$, and $h_0 \sim 7$~km.
Since neutrinos are produced typically between $20 - 30$~km above the surface, the refractive length is not always attained by neutrinos, especially when they travel directly downwards. 

Contrary to solar neutrinos, which also experience an exponentially varying density, the phase evolution of quasi-steriles in the atmosphere is non-adiabatic.
Since the mixing angles are small, the adiabaticity condition is most relevant at the location of the resonance~\cite{Parke:1986jy,Petcov:1987xd,Kuo:1988pn,Kuo:1989qe,Friedland:2000rn}.
There, it takes the form
\begin{equation}
    \gamma \sim \frac{\ell_{\rm osc}}{4 \theta_{S_i}^2} \left|\frac{1}{n_e}\frac{\dd n_{e}}{\dd x}\right| = \frac{\ell_{\rm osc}}{4 \theta_{S_i}^2} \frac{1}{h_0} \sqrt{1  - \sin^2\theta_z\left(\frac{R_0}{R_1}\right)^2},
\end{equation}
where $x$ is the distance traveled by the neutrino, $\ell_{\rm osc} = 4\pi E_\nu / \delta M^2_i$ the vacuum oscillation length, $R_0$ is the Earth's radius, and $R_1 = R_0 + H$ is the production radius of the neutrino, with $H \sim 25$~km being the typical production height.
For downgoing neutrinos with $\cos\theta_z = 1$, $\gamma \sim 2\times10^{3} \times \left( {10^{-2}/\theta_{S_i}}\right)^2$ for $\delta M_i^2 = 230$~eV$^2$, strongly violating the adiabaticity condition ($\gamma \ll 1$).
For the horizon, the change in density is slower, $\dd h/\dd x = 1/80$~km, but not enough to recover adiabaticity, which is only attained in this system for prohibitively large mixing angles.
Finally, we note that oscillations will also be important as neutrinos are produced and detected rather close to the resonance.
In summary, in the quasi-sterile model, the flavor evolution of atmospheric neutrinos is complex and must be modeled as an oscillatory and non-adiabatic system inside an extended source.

\Cref{fig:prob_benchmark} shows the numerical results for the probability of muon neutrino conversion to all invisible quasi-sterile states, $P(\nu_{\mu} \rightarrow {\rm QS})$.
These figures correspond to neutrinos at IceCube coming from the horizontal direction, $\cos(\theta_{z}) = 0$.
The left panel shows a scenario of just one quasi-sterile state $S_2$ mixed with $\nu_2$. 
Our benchmark values, in this case, are
\begin{equation}\label{eq:theta_singleQS}
    \sin{\theta_{S_2}} = 0.01, 
    \,\,
    E_{2}^{\rm res} = 300 \text{ MeV},
\end{equation}
where the resonance is chosen to match the energy region where the MiniBooNE excess is largest, and the remaining parameters are chosen by picking a fixed density $\rho_{\rm MB}$ at MiniBooNE.
By varying the assumption on the MiniBooNE density, we vary the corresponding size of the matter potential and, therefore, the location and size of the resonance in the atmosphere.
For a fixed resonant energy at MiniBooNE and a lower $\rho_{\rm MB}$, we obtain an overall larger disappearance at IceCube thanks to the larger matter potential in the atmosphere and shorter refraction length, which allows the resonance to develop more, albeit at lower energies.

The right panel of \Cref{fig:prob_benchmark} also shows the muon disappearance probability in the full three quasi-sterile neutrino model of \cite{Alves:2022vgn},
\begin{align}\label{eq:theta_threeQS}
    \sin{\theta_{S_1}} = 3.05\times 10^{-3}, \,\, E_{1}^{\rm res} = 290 \text{ MeV},
    \\
        \sin{\theta_{S_2}} = 3.00\times 10^{-3}, \,\,  E_{2}^{\rm res} = 300 \text{ MeV},
    \\
        \sin{\theta_{S_3}} = 2.95\times 10^{-3}, \,\, E_{3}^{\rm res} = 310 \text{ MeV},
\end{align}
where each resonance is induced by a different quasi-sterile neutrino mixing with their respective light-neutrino mass eigenstate.
The remaining parameters, such as the mass splitting, are fixed by requiring the resonant energy above to happen for the MiniBooNE density that is indicated for each curve.
As expected, a greater number of coupled quasi-sterile states results in a larger disappearance of $\nu_{\mu}$.
In the case where all three quasi-sterile states are coupled, we observe a $\sim 10\%$ disappearance in the muon neutrino flux for energies around $600$~GeV.

\begin{figure}[t]
    \centering
    \includegraphics[width=0.49\textwidth]{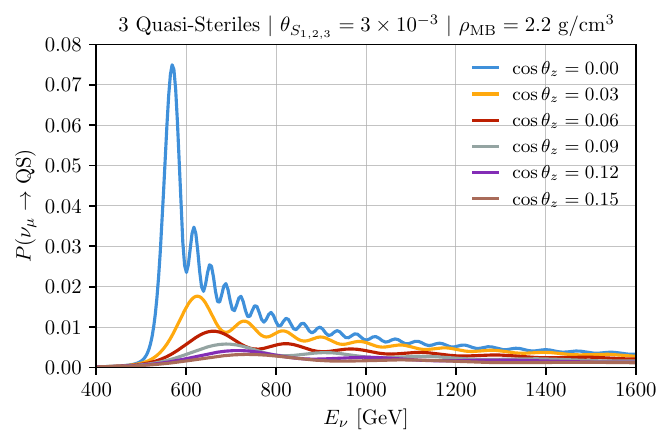}
    \caption{The probability for muon disappearance as a function of the neutrino energy $E_\nu$ for varying zenith angles $\theta_z$ just above the horizon.
    The curves below the horizon are similar.
    We fix $E_{2}^{\rm res} = 300$~MeV and $\rho_{\rm MB} = 2.2$~g/cm$^3$.
    \label{fig:prob_diffzenith}}
\end{figure}

\begin{figure*}[t]
    \centering
    \includegraphics[width=0.49\textwidth]{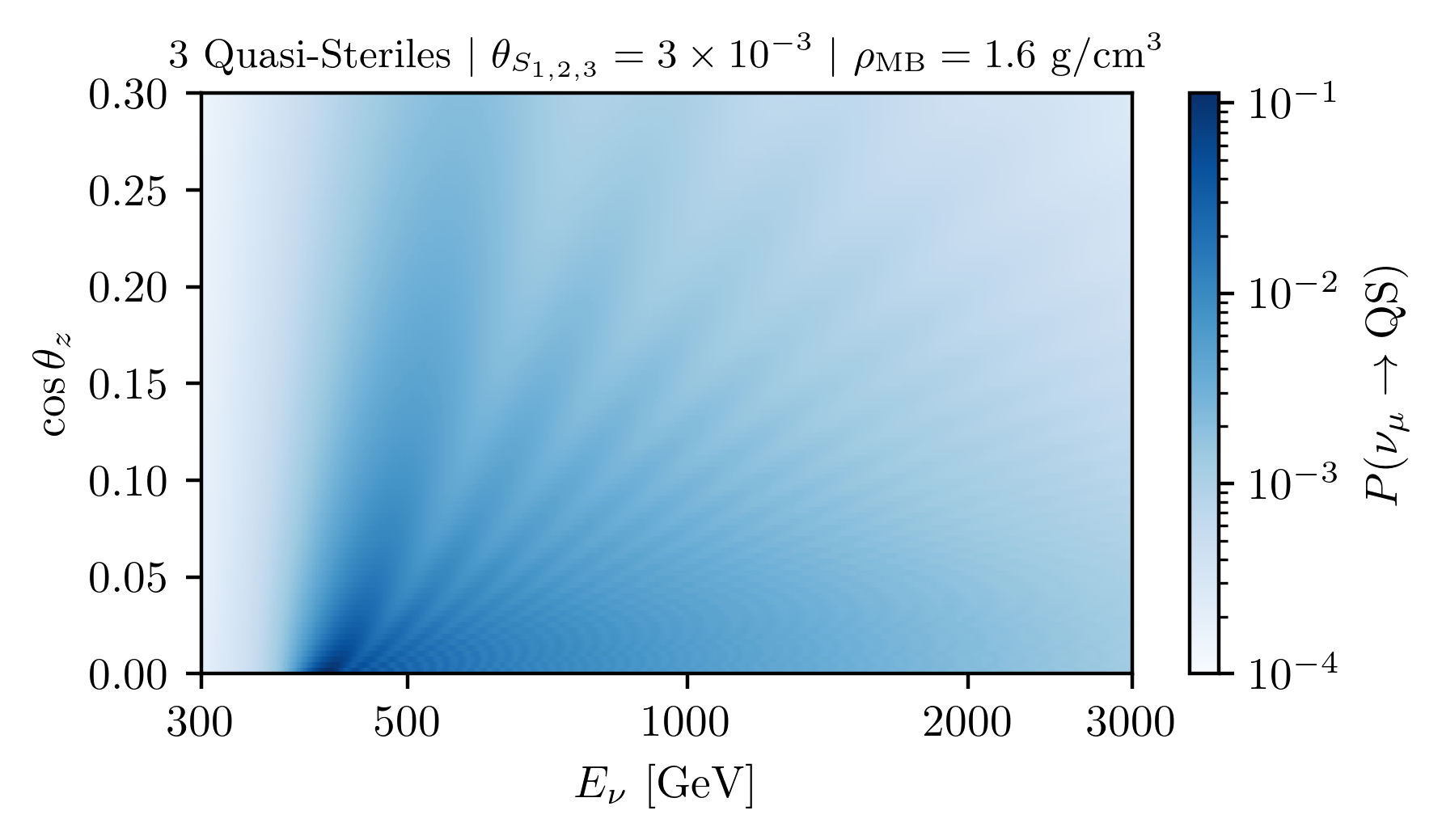}
    \includegraphics[width=0.49\textwidth]{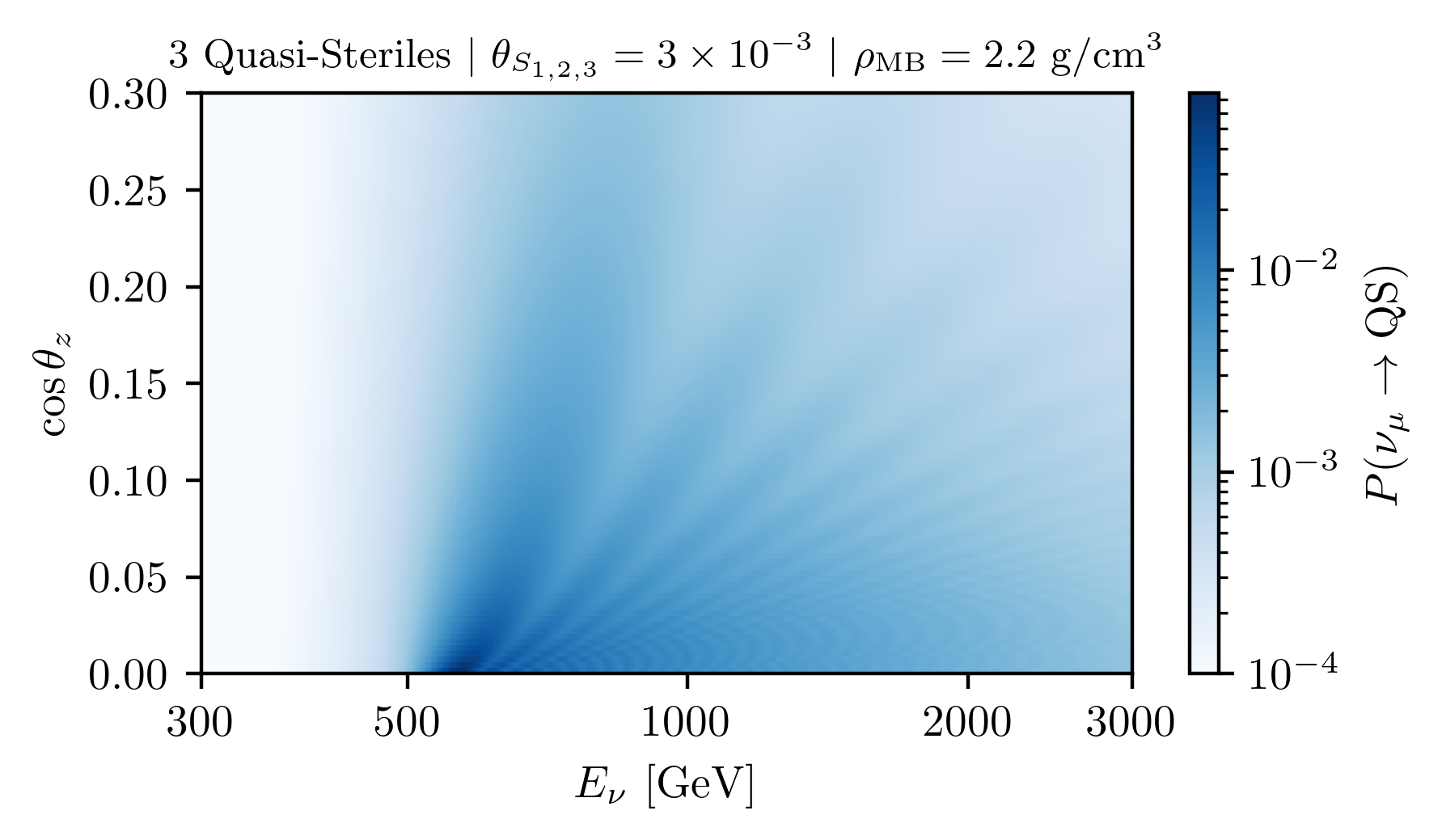}
    \caption{The probability for muon disappearance as a function of the neutrino energy $E_\nu$ and the cosine of the zenith angle $\cos\theta_z$.
    We fix $E_{2}^{\rm res} = 300$~MeV and vary the MiniBooNE soil density from $\rho_{\rm MB} = 1.6$~g/cm$^3$ on the left to $\rho_{\rm MB} = 2.2$~g/cm$^3$ on the right.
    The probability is similar for upgoing events in the same energy and $|\cos\theta_z|$ range.
    \label{fig:prob_2D}}
\end{figure*}

\Cref{fig:prob_diffmixing} illustrates the dependence of the neutrino disappearance, $P(\nu_{\mu} \rightarrow {\rm QS})$, on the mixing between quasi-sterile and active states, $\theta_{S_{1,2,3}}$.
Muon neutrino disappearance becomes almost complete for mixing angles as large as $\theta_{S_{1,2,3}} = 0.05$. 
While such large values of mixing angles are likely ruled out by other neutrino oscillation experiments, they demonstrate the widening of the resonance and the slow approach to the adiabatic limit.
We also note that MiniBooNE is less sensitive to the exact choice of the mixing angle, so that a value of $10^{-2}$, for instance, is still compatible with the excess.
A full exploration of the parameter space of the model, including its compatibility with neutrino data from experiments like T2K~\cite{Smirnov:2021zgn}, as well as solar neutrino experiments, is beyond the scope of our paper.

\Cref{fig:prob_diffzenith,fig:prob_2D} show the probabilities of disappearance for various incoming neutrino energies and directions.
Neutrinos with the same $|\cos\theta_z|$ but opposite signs (above or below the horizon) have virtually the same probability of disappearance in the energy and angular range of interest.
Absorption effects are negligible in the angular region of interest; we did no investigate upgoing events with $\cos{\theta_z} < -0.3$.
The effect of the rapidly varying air density and travel path of neutrinos is visible as a shift and strong suppression of the resonance peak when moving away from the horizon.
In \cref{fig:prob_2D}, we compare two assumptions for the MiniBooNE density.

\subsection{Quasi-steriles at IceCube}

Searches for matter-enhanced resonant neutrino disappearance have been performed over the last ten years in various analysis iterations using IceCube high-energy muon neutrinos~\cite{IceCube:2016rnb,IceCube:2020tka,IceCube:2020phf}.
Though no significant evidence for muon-neutrino disappearance has been found in this analysis, a persistent preference for a light sterile neutrino has increased significance.
The latest iteration of these analyses~\cite{IceCube:2024kel,IceCube:2024uzv} uses a sample of $\lesssim 400,000$ muon neutrinos with a purity of $99.9\%$ contamination of other flavors or atmospheric muons, where a prior iteration --- which has been recently made publicly available~\cite{DVN/9WGYQN_2024} --- had $\sim 300,000$ muon neutrinos showing an improved efficiency of $\sim$~30\% mainly attributed to using Boosted Decision Trees for cut design instead of only straight cuts.
The modeling of the atmospheric neutrino uncertainties, detector effects, and astrophysical neutrino components have also improved significantly over the years needed to take advantage of larger sample sizes.
With the current $\nu_\mu + \overline{\nu}_\mu$ disappearance analyses at IceCube, the uncertainty on the bin-to-bin \emph{uncorrelated} event rate is constrained to approximately $4\%$~\cite{DVN/9WGYQN_2024}.
This value, however, is not representative of the sensitivity of this sample, as the bin-to-bin correlations are important and would strengthen the sensitivity.

The dominant sources of uncertainty are the hadronic interaction models used in the neutrino flux prediction, the properties of the Antarctic ice, and the efficiency of the IceCube sensors.
Improvement in hadronic interaction models is expected to occur with more data from the forward physics program at the LHC.
The optical properties of the Antarctic ice are expected to be significantly better measured over the next years. 
This is due to the expected deployment of a series of calibrations in the ice on the upcoming IceCube-Upgrade, which is expected to be deployed in the 2025-2026 Antarctic summer pole season.
This will significantly reduce systematic errors, especially in the horizontal direction, where neutrinos traverse the largest column density of air before reaching IceCube.

In summary, current IceCube detector efficiencies, event reconstruction capacities, and systematic uncertainties allow for the search for a few percentage-level effects at TeV energies using muon neutrinos.
Upcoming improvements to the IceCube detector array, increased knowledge of forward physics, and machine learning enhancement of event reconstruction promised further improved accuracy. 
Additionally, we expect that KM3NeT, the next-generation kilometer-cube-scaled water neutrino telescope under construction in the Mediterranean Sea, will have a similar capacity for searching for these signatures after sufficient data has been acquired.

\section{Conclusions}
\label{sec:conclusions}

Atmospheric neutrino production takes place across a large extent in the atmosphere.
As neutrinos traverse the low-density air medium, they experience a small matter potential sourced by local protons, neutrons, and electrons.
For the first time, we include these effects in the flavor evolution of neutrinos within the \textsc{nuSQuIDS} package~\cite{Arguelles:2021twb}.
The production height distribution is obtained directly from \textsc{MCEq}, and the varying density is computed according to empirical atmospheric models, where, for instance, air densities go from $\rho\sim 1.3$~mg/cm$^3$ on the Earth's surface to $\rho\sim 0.1$~mg/cm$^3$ at heights of $20$~km, where neutrino production is most frequent.
This effect is a sizeable correction for the standard oscillations of active neutrinos.
While it is already taken into account in low-energy oscillation analyses at Super-Kamiokande~\cite{Super-Kamiokande:2005mbp,Super-Kamiokande:2017yvm} and IceCube~\cite{IceCube:2024xjj}, this is the first time such effects are included in open-source neutrino flavor evolution software.

This new implementation allowed us to study a new physics scenario with resonant flavor conversion in the atmosphere.
The resonance is caused by a large matter potential experienced by \emph{quasi-}sterile neutrinos.
These models have been proposed to explain the excess of low energy events at MiniBooNE by introducing a new matter resonance for neutrinos traveling through the dirt at energies of $E_\nu \simeq 300$~MeV.
If the resonance is induced by a new matter potential sourced by protons, neutrons, or electrons, it would unequivocally predict a corresponding resonant transition in the column density traversed by atmospheric neutrinos of energy $\mathcal{O}(300- 700)$~GeV.
As it so happens, the production of neutrinos from the horizon takes place at a sufficiently high altitude for them to fully develop resonant flavor transitions on their way to the surface of the Earth.
This signature adds to other existing probes of the model, among those cosmology, resonant electron-neutrino disappearance inside the Sun, and other accelerator neutrino experiment data.

At neutrino telescopes, the most promising avenue to search for the new resonance is through searches for a sharp $\nu_\mu \to \nu_\mu$ disappearance at $\cos\theta_z = 0$.
For the benchmark values of the MiniBooNE solutions in Ref.~\cite{Alves:2022vgn}, we find that the resulting $\nu_\mu$ disappearance at IceCube in the direction of the horizon ranges from $4\%$ to $10\%$, depending on the assumed density and resonant energy at MiniBooNE.
These values quickly increase when considering larger but still sub-percent-level mixing angles, which are also consistent with the MiniBooNE excess.
They also exceed the uncorrelated bin-to-bin uncertainties of the latest IceCube measurements~\cite{IceCube:2024kel,IceCube:2024uzv}, suggesting that full analysis that takes into account energy and angular correlations can meaningfully constrain the existence of such a resonance.
With an improved energy resolution, we also expect KM3NeT to be sensitive to such effects.
In addition to the Earth-induced resonance used to search for eV-sterile neutrinos, neutrino telescopes can once again weigh in on the short-baseline puzzle, taking advantage of a different energy range and different matter background density.

\begin{acknowledgments}
We thank Alex Wen for discussions on the uncertainties of atmospheric neutrinos at IceCube.
The authors dedicate this article to the memory of Tom Gaisser, whose impactful contributions in our field echo in this work~\cite{Halzen:2024usa}.
The work of MH is supported by the Neutrino Theory Network Program Grant \#DE-AC02-07CHI11359 and the US DOE Award \#DE-SC0020250.
CAA are supported by the Faculty of Arts and Sciences of Harvard University, the National Science Foundation (NSF), the NSF AI Institute for Artificial Intelligence and Fundamental Interactions, the Research Corporation for Science Advancement, and the David \& Lucile Packard Foundation.
CAA and IMS were supported by the Alfred P. Sloan Foundation for part of this work.
CS was supported by the \textit{Harvard College Research Program} (HCRP) and funding from the Faculty of Arts and Sciences of Harvard University through this work.
IMS was funded by the grant ST/T001011/1.
\end{acknowledgments}

\bibliographystyle{apsrev4-1}
\bibliography{main}{}

\end{document}